
\input phyzzx
\input box


\def\k{$\kappa$-}

\pubnum={57/93}
\date{November 1993}
\titlepage
\title{Classical and Quantum Mechanics of Free \k Relativistic Systems}
\author{
Jerzy Lukierski$\sp{\ddagger}$\foot{On leave of absence from the
Institute of Theoretical Physics, University of Wroclaw, pl. M. Borna 9,
 50-205 Wroclaw, Poland}\foot{Supported by EC grant 3510PL922519}\break
Henri Ruegg$\sp{\sharp}$\foot{Partially supported by the Swiss National Science
Foundation, OFES contract 93.0083 and Human Capital and Mobility
EC contract ERBCHRXCT920069}\break
  Wojtek J. Zakrzewski $\sp{\ddagger}$     }
\address{
$\sp{\ddagger}$Department of Mathematical Sciences\break
      University of Durham, South Road, Durham DH1 3LE, England\break\break
$\sharp$
D\'epartement de Physique Th\'eorique, Universit\'e de Gen\`eve, 24 quai
Ernest-Ansermet,
1211 Gen\`eve 4, Switzerland\break\break
\break}
\vskip -1cm
\abstract{
We consider the Hamiltonian and Lagrangian formalism describing free
\k-relativistic particles with their four-momenta constrained to the
\k-deformed mass shell. We study the modifications of the formalism which
follow from the
introduction of space coordinates  with nonvanishing Poisson brackets
and from the
redefinitions of the energy operator. The quantum mechanics of free
\k-relativistic particles
and of the free \k-relativistic oscillator is also presented.
It is shown that the \k-relativistic oscillator describes a quantum
statistical ensemble with finite Hagedorn temperature. The relation to a
\k-deformed
Schr\"odinger quantum mechanics in which the time derivative
 is replaced by  a finite
 difference derivative is also discussed. }

\endpage
\chapter{Introduction}

\REF\rone{V.G.
Drinfeld - {\it Proc of XX International Math. Congress, Berkeley} Vol.I
, 798 (1986)}

\REF\rtwo{L.D. Faddeev, N.Yu. Reshetikin and L.A. Takhtajan -
 {\it Algebra and Analysis} {\bf 1}, 178 (1989)}

\REF\rthree{S.L. Woronowicz - {\it Comm. Math. Phys}
 {\bf 111}, 613 (1987), ibid.  {\bf 122}, 125 (1989)}

\REF\rfour{J. Lukierski, A. Nowicki, H. Ruegg and V.N. Tolstoy - {\it Phys.
Lett.} {\bf B 264}, 331 (1991)}

\REF\rfive{O. Ogievetsky, W.B. Schmidke, J. Wess
and B. Zumino - {\it Comm. Math. Phys.} {\bf 150}, 495 (1992)}

\REF\rsix{J. Lukierski and A. Nowicki - {\it Phys. Lett.} {\bf B 279}, 299
(1992)}

\REF\rseven{S. Giller, J. Kunz, P. Kosinski, M. Majewski
 and P. Maslanka - {\it Phys. Lett.} {\bf B 286}, 57 (1992)}

\REF\reight{J. Lukierski, A. Nowicki and  H. Ruegg -
{\it Phys. Lett.} {\bf B 293}, 344 (1992)}

\REF\rnine{J. Lukierski, H. Ruegg and W. Ruhl - {\it Phys. Lett.} {\bf B 313 },
  357(1993)}

\REF\rten{H. Bacry - {\it Journ. Phys} {\bf 26A}, 5413 (1993)    }

\REF\releven{M. Schlicher, W. Weich and R. Weixler - {\it Zeit. f. Phys.}
{\bf C 53}, 79 (1992)}

\REF\rtwelve{S. Majid - {\it Jour. Math. Phys.}  {\bf 34}, 2045 (1993)     }

\REF\rthirteen{M. Chaichan and A. Demitchev - {\it Phys. Lett.}
 {\bf B304}, 220 (1993); see also: M. Chaichan and A. Demitchev -
 Helsinki Univ. preprint HU-TFT-93-24 }

\REF\rfourteen{S. Zakrzewski - Quantum Poincar\'e group related to
$\kappa$-Poincar\'e algebra, Warsaw Univ. preprint, (March 1993)}

\REF\rfifteen{ J. Lukierski and H. Ruegg -
Geneve University preprint UGVA-DPT 1993/
07-825 (19993)}

\REF\rsixteen{L. Woronowicz - Lecture at I-st Carribean  Spring  Spring School,
Guadeloupe
(May-June 1993)}

\REF\rseventeen{J. Lukierski, A. Nowicki and  H. Ruegg - Proceedings of
``Symmetry VI", Bregenz, August 1992, ed. B. Gruber, {\it Plenum Press} (1993)}

\REF\reighteen{H. Bacry - {\it Phys. Lett} {\bf 306 B}, 41, (1993)}

\REF\rnineteen{M. Maggiore - Pisa University preprints IFUP-TH 19/93
and IFUP-TH 38/93 (1993)}

\REF\rtwentyonea{R. Hagedorn - {\it Nuovo Cim. Suppl.} {\bf 3}, 147 (1965)}

\REF\rtwentyoneb{H. Satz - {\it Phys. Rev.} {\bf D5}, 3231 (1979)}

\REF\rtwenty{P. Maslanka - Lodz University preprint HEP-TH9310027 (1993)}

\REF\rtwentyone{H. Bacry - {\it Ann. Inst. H. Poincar\'e}, {\bf 49}, 245
(1988)}

\REF\rtwentytwo{J. Lukierski, A. Nowicki and H. Ruegg - {\it Journ. of Geometry
and Physics}, {\bf 11}, 425   (1993)}

\REF\rtwentythree{J. Lukierski, A. Nowicki and  H. Ruegg - Geneva University
preprint  UGVA-DPT 1993/3-812(1993), {\it J. Math. Phys.} in print}

\REF\rtwentyfour{C. Quigg and J.L. Rosner - {\it Physics Reports}, {\bf 56},
167
(1979)}

\REF\rnewone{see \eg\ M. Abramowitz and I. Stegun - Handbook of Mathematical
Functions, 804, Nat. Bureau of Standards,  (1964)}

\REF\rtwentysix{P. Caldirola - {\it Lett. Nuovo Cimento} {\bf 16}, 151 (1976);
ibid {\bf 18}, 465 (1978)}

\REF\rtwentyseven{P. Caldirola and E. Montaldi - {\it Nuovo Cimento} {\bf 84B},
27 (1984); ibid {\bf 90B}, 58 (1985)}

\REF\rtwentyeight{A. Janussis - {\it Nuovo Cimento} {\bf 84B}, 27 (1984); ibid
{\bf 90B}, 58 (1985)}

\REF\rtwentyfive{M. Klimek, J. Lukierski and W.J. Zakrzewski, in preparation}

\REF\rtwentynine{C. Fronsdal and A. Galindo - {\it Lett. Math. Phys.} {\bf 27},
59 (1993)}

\REF\rthirty{F. Bonechi, E. Celeghini, R. Giacchetti, C.M. Pere\~na, E. Sorace
and M. Tarlini - Firence University preprint DFF192/9/93 (1993)}

Recently, many proposals were presented showing how to apply the ideas of
quantum deformations [\rone-\rthree] to the $D=4$ Poincar\'e algebra
[\rfour-\rten] as well as
the $D=4$ Poincar\'e group [\releven-\rsixteen].
The $\kappa$-deformation of the $D=4$ Poincar\'e algebra, first proposed
in [\rfour,\reight], leads to the  modification of relativistic symmetries with
the
three-dimensional $E(3)$ subalgebra unchanged. The deformation
parameter $\kappa$ describes the fundamental mass in the theory
and the limit $\kappa\rightarrow \infty$ corresponds to the undeformed case.

As a consequence of the \k deformation the mass shell condition is changed as
 follows ( $\vec p\sp2\equiv p_1\sp2+p_2\sp2+p_3\sp2$) [\reight]:

$$p_0\sp2- \vec p\sp2=m\sp2\quad\rightarrow\quad \bigl(2\kappa sinh{p_0\over
2\kappa}\bigr)\sp2-\vec p\sp2=m\sp2.\eqn\eonethree$$
The \k-deformed mass-shell condition leads to the following modification of
 the Hamiltonian of  free
\k-relativistic particles:
$$H=P_0=\sqrt{\vec P\sp2+M\sp2}\;
\rightarrow\;H_0\sp{\kappa}=P_0=2\kappa\,arcsinh{\sqrt{\vec P\sp2+M\sp2}\over
2\kappa}.\eqn\eonetwo$$

In the \k-deformed Poincar\'e algebra, which we describe in section 2,
 the four-momenta commute and
we can introduce the space-time
dependence by
the standard Fourier transforms of the four-momenta functions. In such a way we
can choose
 the \k-relativistic classical mechanics
to be formulated in terms of
 space coordinates satisfying standard relativistic Poisson brackets:
$$\{x\sp{\mu},p\sp{\nu}\}=i\eta\sp{\mu\nu},\qquad \hbox{diag}\eta=(-1,1,1,1).
\eqn\eonefour$$
The form of the  Hamilton equations of motion remains also unchanged \ie\
we have
$$\dot x_i={\partial H\sp{\kappa}\over \partial p_i},\quad
 \dot p_i=-{\partial H\sp{\kappa}\over \partial x_i}.\eqn\eonefive$$
In such an approach  the only change due to the \k-deformation
 is the explicit form of the
Hamiltonian (see \eonetwo\ for the free case).

In section 3 we present the Hamiltonian and Lagrangian formalism
for both the massive and massless
free \k-relativistic particles described by the Hamiltonian
 $H_0\sp{\kappa}$.
As is known [\rseventeen,\reighteen,\rten], however, the velocity decreases
with the increase
of energy  at large
energies ($E\gg \kappa m$) in such a model.

 To eliminate the velocity decrease with the increase of energy
we present in section 4 two
possible remedies:
\item{i)} The introduction of space coordinates with nonvanishing Poisson
brackets.
We consider massive spinless systems and we
introduce the three space coordinates $x\sp{i}$  as the
functions of the \k deformed Poincar\'e algebra generators.
 The standard choice, with  space variables satisfying \eonefour\ (for
$\mu,\nu=1,2,3$) was found by
 Bacry [\reighteen, \rten]. Recently Maggiore [\rnineteen]
 proposed other functions
of the \k Poincar\'e generators
which even for a spinless system describe space coordinates with
nonvanishing Poisson brackets. It appears that such a proposal fits nicely
into the schemes in which the symplectic tensor determining
the Hamilton equations is not canonical.

\item{ii)} The definition of the physical energy $\tilde P_0$ of
 the \k-deformed
system by
$$ \tilde P_0=2\kappa \,sinh{P_0\over 2\kappa}.\eqn\eoneoneone$$
In this case the \k-deformation does not affect the mass-shell
condition but modifies the classical additivity of the energy. At the end of
section 4 we show
 how these
two modifications are related.

In section 5 we describe the Schr\"odinger equation with the hamiltonian
$H\sp{\kappa}$
which we shall call the \k-relativistic Schr\"odinger equation.
We  consider two examples:  a free \k-relativistic particle and
 a free \k-relativistic harmonic oscillator.
For the \k-relativistic oscillator we calculate its partition
function and show that the quanta of free \k-oscillators
exist as a statistical ensemble only below a critical temperature $T_c$
which is usually called the Hagedorn temperature [20,21].
We also show that the quantum \k-relativistic systems
 can be described by the Schr\"odinger equation in which the time derivative is
replaced
by a finite difference time derivative. We call
this equation the \k-deformed Schr\"odinger equation.

In this paper we do not take into consideration the nontrivial coproduct
structure of the four-momenta  generators of the \k-deformed quantum Poincar\'e
algebra, which is a Hopf algebra.
The noncommuting space-time coordinates
can, however, be introduced as the generators of the quantum
\k-Poincar\'e group dual to the \k-deformed Poincar\'e algebra [14,15].
 Because the  quantum Poincar\'e group translations $\hat x_{\mu}$ describing
space coordinates
are dual to the noncocommutative four-momenta  they must be noncommuting
and  satisfy the algebra [\rfourteen, \rfifteen, 22]:
$$\eqalign{[\hat x_i,\,\hat x_j]&=0\cr
[\hat x_0,\,\hat x_j]&={1\over \kappa}\hat x_j.\cr}\eqn\eonetwotwothree$$

\noindent In the last section of this paper we indicate what emerges if we
consider
the full Hopf algebra structure of the \k-deformed Poincar\'e algebra.
We believe that such an  approach is very promising but, at present,
we are only able to mention some aspects of a formulation
with noncommutative space coordinates as well as non-additive four-momenta.

\chapter{\k-deformed quantum Poincar\'e algebra}
First we recall here the basic formulae describing the quantum deformation of
the $D=4$ Poincar\'e algebra
with a fundamental mass-like parameter $\kappa$ [\rfour,\rseven-\rten].
The quantum deformations are described by a noncocommutative Hopf algebra with
its algebra and coalgebra sectors and with the notions
of  antipodes and a counit [\rone-\rthree].
The \k deformation of the Poincar\'e algebra is given by the formulae
(with \k  real mass-like parameter - see [\reight]):
\item{a)} algebra sector
$$\eqalign{[M_i, M_j]=&i\epsilon_{ijk}M_k,\qquad [P_{\mu},P_{\nu}]=0\cr
[N_i, M_j]=&i\epsilon_{ijk}N_k\cr
[M_i, P_j]=&i\epsilon_{ijk}P_k,\qquad [M_{i},P_{0}]=0\cr
[N_i, P_j]=&i\kappa \delta_{ij}\,sinh{P_0\over \kappa}\cr
[N_i, P_0]=&iP_j\cr
[N_i, N_j]=&-i\epsilon_{ijk}(M_kcosh{P_0\over \kappa}-{1\over 4\kappa\sp2}P_k
(\vec P\vec M))\cr}\eqn\etwoone$$
\item{b)} coalgebra
$$\eqalign{\Delta(M_i)=&M_i\otimes I+I\otimes M_i\cr
\Delta(N_i)=&N_i\otimes exp({P_0\over 2\kappa})
+exp(-{P_0\over 2\kappa})\otimes N_i\cr
+{1\over 2\kappa}\epsilon_{ijk}\bigl(&P_j\otimes M_kexp({P_0\over 2\kappa})
+exp(-{P_0\over 2\kappa})M_j\otimes P_k\bigr)\cr
\Delta(P_i)=&P_i\otimes exp({P_0\over 2\kappa})
+exp(-{P_0\over 2\kappa})\otimes M_i\cr
\Delta(P_0)=&P_0\otimes I+I\otimes P_0\cr}\eqn\etwotwo$$
\item{c)} antipodes
$$\eqalign{S(M_i)=&-M_i,\qquad S(P_{\mu})=-P_{\mu},\cr
S(N_i)=&-N_i+{3i\over 2\kappa} P_i.\cr}\eqn\etwothree$$

\noindent The bilinear mass square Casimir of the
 Poincar\'e algebra is deformed as follows
$$C_1\,\rightarrow\,C_1=\vec P\sp2+2\kappa\sp2(1-cosh{P_0\over \kappa})=\vec
P\sp2 -
(2\kappa sinh{P_0\over 2\kappa})\sp2,\eqn\etwofour$$
where the eigenvalue $C_1=-M\sp2$, determines the \k relativistic rest mass
(compare with \eonethree).

The \k Poincar\'e algebra can be realised by vector fields on a commuting
four-momentum space. The realisation on scalar fields $\phi(p_{\mu})$
can be written as follows:

$$\eqalign{P_{\mu}=p_{\mu},\qquad& M_{ij}=-i(p_i\partial_j-p_j\partial_i),\cr
N_i=M_{i0}=&-i(p_i\partial_0-\kappa sinh{p_0\over
\kappa}\partial_i).\cr}\eqn\etwofive$$
In this representation the variables $\vec p$ and $p_0$ are treated as
independent variables (and so we have an off-shell representation).
On the \k-relativistic mass shell  we have $C_1=-M\sp2$
and the realisation of the \k-Poincar\'e algebra
on the three-dimensional momentum space is obtained by the following
modification of $N_i$ in \etwofive:
$$N_i=i\omega \sqrt{1+{\omega\sp2\over 4\kappa\sp2}}{\partial \over \partial
p_i},\eqn\etwoseven$$
where ${\omega\over 2\kappa}=sinh{p_0\over 2\kappa}$. But as ${\omega\over
2\kappa}\sqrt{1+{\omega\sp2\over 4\kappa\sp2}}=
sinh{p_0\over 2\kappa}cosh{p_0\over 2\kappa}$ = ${1\over 2}sinh{p_0\over
\kappa}$
we see that on the \k-relativistic mass shell we recover Bacry's position
operator [\reighteen, \rten]
$$X_i\sp{B}={1\over 2}\Bigl\{{1\over \kappa sinh{P_o\over \kappa}},\,N_i\Bigr\}
\quad\leftrightarrow\quad X_i\sp{B}=i{\partial \over \partial
p_{i}}.\eqn\etwoeight$$

Recently, Maggiore  [\rnineteen] has proposed another form of the position
operator
$$X_i\sp{M}=i\sqrt{1+{\omega\sp2\over 4\kappa\sp2}}{\partial \over \partial
p_i}
\eqn\etwonine$$
or, equivalently,
$$X_i\sp{M}={1\over 2}\Bigl\{{1\over 2\kappa sinh{P_0\over
2\kappa}},\,N_i\Bigr\}.\eqn\etwoten$$
It is easy to check that
$$\eqalign{[X_i\sp{M},X_j\sp{M}]&={1\over 4\kappa\sp2}M_{ij},\cr
[X_i,P_j]=-i&\delta_{ij}cosh{P_0\over 2\kappa}.\cr}\eqn\etwoeleven$$
The advantage of the formula \etwonine\  stems from the fact that it reproduces
the conventional position operator for relativistic theories [23]
$$X_i\sp{M}={1\over 2} \bigl\{{1\over \tilde
P_0},\,N_i\bigr\},\eqn\etwoeleven$$
where
$$\tilde P_0=2\kappa sinh{P_0\over 2\kappa}=P_0+{1\over
24\kappa\sp2}P_0\sp3\,+\,O({1\over \kappa\sp4}).\eqn\etwoelevena$$

The new choice \etwoelevena\ of the energy operator
leads to the following new form of the last three relations in \etwoone
$$\eqalign{[N_i, P_j]=&i \delta_{ij}\,\tilde P_0\sqrt{1+{\tilde P_0\over
4\kappa}},\cr
[N_i,\tilde P_0]=&i P_j
 \sqrt{1+{\tilde P_0\over 4\kappa}},\cr
[N_i, N_j]=&-i\epsilon_{ijk}\Biggl(M_k\sqrt{1+{\tilde P_0\sp2\over \kappa\sp2}
\Bigl(1+{\tilde P_0\sp2\over 4\kappa\sp2}\Bigr)}
-{1\over 4\kappa\sp2}P_k
(\vec P\vec M)\Biggr).\cr}\eqn\etwotwelve$$
In the coproducts $\Delta(N_i),\, \Delta(P_i)$ (see \etwotwo)
we have to make the replacement
$$exp\{{{\pm P_0\over 2\kappa}}\}\,\rightarrow\,{\tilde P_0\over 2\kappa}\pm
\sqrt{1+{\tilde P_0\sp2\over 4\kappa\sp2}}.\eqn\etowthirteen$$
Furthermore, we find (see also [\rtwentytwo]) that
$$\Delta(\tilde P_0)= \tilde P_0\otimes \sqrt{1+{\tilde P_0\over 4\kappa\sp2}}
+\sqrt{1+{\tilde P_0\over 4\kappa\sp2}}\otimes \tilde P_0
\eqn\etwofourteen$$
and  $$C_1=\vec P\sp2-\tilde P_0\sp2=-M_0\sp2.\eqn\etwofifteen$$

The freedom of defining the energy operator has recently been pointed out
in [\rten, \rtwentythree]. It appears that for the choice \etwoelevena\
of the energy variable one obtains the conventional mass-shell
condition and the usual formula for the energy, namely,
$$\tilde P_0=\sqrt{\vec P\sp2 +M_0\sp2}.\eqn\etwosixteen$$

The relation (2.12) between  $P_0$ and $\tilde P_0$
 remains also valid
 if we consider the energy of a composite system. Using the
respective coproduct formulae we obtain
($\omega_i=\sqrt{\vec p_i\sp2+M_i}$) in the $P_0$ description
$$p_0\sp{(1+2)}=p_0\sp2+p_0\sp2=arcsinh{\omega_1\over
2\kappa}+arcsinh{\omega_2\over 2\kappa}=arcsinh{\omega_{12}\over
2\kappa},\eqn\etwoseventeena$$
where
$$\omega_{12}=\omega_1\sqrt{1+{\omega_2\sp2\over
4\kappa\sp2}}+\omega_2\sqrt{1+{\omega_1\sp2\over
4\kappa\sp2}},\eqn\etwonineteen$$
 while in the $\tilde P_0$ description we obtain
$$\tilde p_0\sp{(1+2)}=\omega_{12}=2\kappa\,sinh{p_0\sp{(1+2)}\over 2\kappa}.
  \eqn\etwoseventeenb$$

In the next section we describe the free  \k-relativistic  particles
using the $P_0$ picture with the conventional \k-Poincar\'e algebra (2.1-3),
additive energy and commuting space coordinates.
The case of noncommuting coordinates as well as of the \k-deformed
theory with  the choice \etwofifteen\ of the undeformed mass-shell
condition will be considered in section 4.

\chapter{\k-relativistic classical mechanics - the case with commuting space
coordinates}
In this section we consider standard Hamiltonian mechanics with
the Hamiltonian defined by the \k-relativistic mass shell condition
(see \eonetwo\ and \etwofour).
\item{a)} Hamiltonian formalism

The equations of motion for the \k-relativistic particle take the form
$$\dot x_i={\partial H_0\sp{\kappa}\over \partial p_i}={1\over
\sqrt{1+{\omega\sp2\over 4\kappa\sp2}}}{p_i\over \omega}\eqn\efourone$$
$$\dot p_i=0,\eqn\efourtwo$$
where $\omega=(\vec p\sp2+M\sp2)\sp{1\over 2}$.
 The formula \efourone\ describes
a modification of the Einstein velocity formula $v_i={p_i\over \omega}$
and has been discussed previously [17,18].
 From \efourone\ we obtain ($v_i\equiv \dot x_i$, $v\sp2=v_i v_i$)
$$v\sp2={\vec p\sp2\over (1+{\vec p\sp2+M\sp2\over 4\kappa\sp2})(\vec
p\sp2+M\sp2)}\eqn\efourthree$$
and we find that for $$\vec p\sp2=\vec
p_{max}\sp2=M(4\kappa\sp2+M\sp2)\sp{1\over 2},\eqn\efourfour$$
the maximal velocity $v_{max}\sp2$ is given by
$$v_{max}\sp2={4\kappa\sp2\over (M+\sqrt{M\sp2+4\kappa\sp2})\sp2}=1-{M\over
\kappa}
+O\Bigl({M\sp2\over \kappa\sp2}\Bigr).\eqn\efourfive$$
 From  \efourthree\ we see that
\item{a)} if $\vec p\sp2<\vec p_{max}\sp2$ we are in the {\bf conventional}
(nonrelativistic or relativistic)
{\bf regime}, with the velocity increasing with energy $p_0$ (we recall
 that in the conventional theory the energy
$p_0$ is a monotonic function of $\vec p\sp2$; see \eonetwo).
\item{b)} if $\vec p\sp2>\vec p_{max}\sp2$ we are in the \k-relativistic
regime, with the velocity {\bf diminishing} with energy.

The double-valuedness of energy (for $M>0$) as a function of momenta implies
that one has two different Lagrangians - for $v\sp2<v_{max}\sp2$ (we shall
call this the conventional Lagrangian $L\sp{c}$) and for $v\sp2>v_{max}\sp2$
(the
nonconventional Lagrangian, which we shall denote by $L\sp{nc}$).
For $M=0$ (scalar \k-photons) we have only the nonconventional Lagrangian,
because $v_{max}\sp2\vert_{M=0}=1.$

\item{b)} Lagrangian formalism.
To obtain  the Lagrangian we proceed  as follows:
\item{i)}
First we invert the velocity formula \efourone\
$$p_i=f(v\sp2,M\sp2)v_i.\eqn\efoursix$$
\item{ii)} Then as
$$p_i={\partial  L(v\sp2)\over \partial v_i}=2v_iL'(v\sp2),\eqn\efourseven$$
we see that the Lagrangian is obtained by comparing \efoursix\ and \efourseven,
\ie\ it is given by
$$L(v\sp2)={1\over 2}\int dv\sp2f(v\sp2,M\sp2).\eqn\efoureight$$

We shall consider separately the massless and the massive cases.

\item{b1)} A classical mechanics of scalar \k-photons.

\noindent By putting $M=0$ in \efourone\ we get
$$p_{i}={2\kappa\over v\sp2}\sqrt{1-v\sp2}v_i\eqn\efournine$$
or
$$L_{M=0}\sp{NC}=\kappa\int dv\sp2{\sqrt{1-v\sp2}\over v\sp2}=
2\kappa\Bigl\{\sqrt{1-v\sp2}+{1\over 2}ln{(1-(1-v\sp2)\sp{1\over 2}\over
(1+(1-v\sp2)\sp{1\over 2}}\Bigr\}.\eqn\efourten$$
It is easy to check the consistency of \efourten\ with the known expression for
the \k-relativistic Hamiltonian for $M=0$ (see \eonetwo). Because $M=0$,
$$\vec p\vec v=2\kappa \sqrt{1-v\sp2}\eqn\efoureleven$$
 and we find that
$$\eqalign{H=&\vec p\vec v-L_{M=0}\sp{NC}=\kappa ln{(1-(1-v\sp2)\sp{1\over
2})\over (1+(1-v\sp2)\sp{1\over 2})}\cr
=2\kappa &arcsinh\Bigl({(1-v\sp2)\sp{1\over 2}\over v}\Bigr)=2\kappa
\;arcsinh\Bigl({\vert \vec p\vert\over 2\kappa}\Bigr).\cr}\eqn\efourelevenb$$
In order to understand \efournine\ in terms of Einsteinian relativistic
mechanics let us write
$$p_i={m_0(v\sp2)v_i\over \sqrt{1-v\sp2}}.\eqn\efourtwelve$$
We see that for $\kappa<\infty$ the ``rest-mass" of the photon is velocity (or
three-momentum) dependent
$$m_0(v\sp2)=2\kappa ({1\over v\sp2}-1)={\vec p\sp2\over
2\kappa}.\eqn\efourthirteen$$
It is clear that as $\kappa\rightarrow \infty$ we find that $m_0\rightarrow 0$.
\item{b2)}  A classical mechanics of massive \k relativistic particles.

\noindent
Inverting \efourone\ for $M\ne0$ gives us two solutions for $\omega\sp2=\vec
p\sp2+M\sp2$. They are
$$\omega_{\pm}\sp2={2\kappa\sp2\over v\sp2}\Bigl[1-v\sp2\pm \sqrt{(1-v\sp2)\sp2
-{v\sp2M\sp2\over \kappa\sp2}}\Bigr].\eqn\efourfourteen$$
Of these
$$\omega_-\sp2\rightarrow {M\sp2\over (1-v\sp2)}\qquad \hbox{as}\quad
\kappa\rightarrow \infty,\eqn\efourfifteen$$

\noindent while
$$\omega_+\sp2\rightarrow\infty,\eqn\efourfifteena$$
thus showing that in the conventional region ($\vec p\sp2<\vec p_{max}\sp2$)
we should use $\omega\sp2=\omega_-\sp2$, while in the
\k relativistic one ($\vec p\sp2>\vec p_{max}\sp2$) we take
$\omega\sp2=\omega_+\sp2$.
Of course $\omega_+\sp2=\omega_-\sp2$ at $v\sp2=v_{max}\sp2$.

To calculate $L$ in both regimes we observe that \efourthree\ implies
that
$$\vec p\vec v=v\sp2\omega \sqrt{1+{\omega\sp2\over
4\kappa\sp2}},\eqn\efoursixteen$$
and so assuming that
$$L=\vec p\vec v-H,\eqn\efourseventeen$$
where
$$H=p_0=2\kappa\sp2\,arcsinh{\omega\over 2\kappa}, \eqn\efoureightteen$$
we find that $L$ is given by
$$L_{\pm}=v\sp2\omega_{\pm}\sqrt{1+{\omega_{\pm}\sp2\over
4\kappa\sp2}}-2\kappa\,arcsinh{\omega_{\pm}\over 2\kappa}.\eqn\efournineteen$$
It is easy to check that as $\kappa\rightarrow \infty$
$$L_+\rightarrow \infty, \qquad L_-\rightarrow
-M\sqrt{1-v\sp2}.\eqn\efourtwenty$$

It is interesting to calculate the nonrelativistic ($v\sp2\rightarrow 0$) limit
of both Lagrangians. The calculations are somewhat involved but as
$$ \omega_-\sim M\Bigl(1+{v\sp2\over 2}\Bigr)+{v\sp2M\over 2}\alpha\sp2
\,+\,O(v\sp4),
\eqn\efourtwentyone$$
where $\alpha={M\over 2\kappa}$, and
$$\omega_+\sim {2\kappa\over v}\,+\,O(v),\eqn\fourtwentytwo$$
we find that
$$L_-\sim\hbox{const}\,+\,Mv\sp2\Bigl[1-{1+\alpha\sp2+\alpha\sqrt{1+\alpha\sp2}\over 2(\alpha+\sqrt{1+\alpha\sp2})}\Bigr]\,+\,O(v\sp4)\eqn\efourtwentythree$$
and
$$L_+\sim\,2\kappa\,ln\,v\,+\,\hbox{const}\,+\,O(lnv\,v\sp2).\eqn\efourtwentyfour$$
Thus we see that $L_-$ has a conventional nonrelativistic limit, with the mass
renormalised by a function $f(\alpha)$, where $\alpha={M\over 2\kappa}$.
The Lagrangian for a particle in the \k-relativistic regime is somewhat
unconventional and, clearly, deserves further study.

\chapter{\k-relativistic classical mechanics: \k-deformation of the
standard symplectic structure}
The formalism of quantum ${\kappa}$-Poincar\'e algebra naturally introduces the
four-momentum
variables. In order to have the complete \k-deformed kinematical scheme we have
to supplement the momenta operators with the position
operators $X_{\mu}$. The algebra
consisting of the generators ($P_{\mu},\,M_{\mu\nu},\,X_{\mu}$) will be
called the \k-deformed relativistic kinematic algebra. Such an algebra
describes the kinematics of \k-relativistic theories.
If we restrict our interest to three-dimensional theories the corresponding
algebra will be formed out of the generators $(P_i,\,M_{ij},\,P_{0},\,X_i$),
with $P_0$ (the energy operator) commuting with all other generators. Such an
algebra
will be called the three-dimensional kinematic algebra.

The formulae \etwofive\ show that in the spinless case
 ($\vec P\cdot \vec M=0$) we can embed the \k-deformed relativistic kinematic
algebra in the enveloping algebra of the Heisenberg algebra
 with commuting coordinates. On the other hand  the formulae \etwoeight\
and \etwoten\
show that the generators $X_i$ can be expressed in terms of \k-Poincar\'e
generators, \ie\ the three-dimensional kinematic algebra can be embedded into
$U_{\kappa}(P_4)$.
\par

There exists a freedom in the definition of the position
variables in the kinematic algebra. This freedom,
on the classical level, comes from  the possibility
of  modifying of the canonical commutation relations
\eonethree\ (for $\mu,\nu=1,2,3$) by introducing a nonstandard symplectic
structure. In fact, the Hamiltonian formalism is determined uniquely
if we specify
\item{a)} the Hamiltonian $H(x\sp{i},p\sp{i})$
\item{b)} symplectic tensor $\omega\sp{AB}$, which defines the Hamilton
equations of motion ($(x\sp{i},p\sp{i})\equiv Y\sp{i})$
$${\partial Y\sp{A}\over \partial t}=\omega\sp{AB}{\partial H\over
\partial Y\sp{B}},\eqn\eoneeight$$

\noindent where the tensor $\omega\sp{AB}$ is determined by a nondegenerate
closed two-form $\omega_2=\omega_{AB}dY\sp{A}\wedge dY\sp{B}$, \ie\
$$d\omega_2=0,\quad \omega\sp{AB}\omega_{BC}=\delta\sp{A}_C.\eqn\eonenine$$
In our case we make the following choice of the Poisson brackets
consistent with the \k-deformation of the Poincar\'e algebra ($i=1,2,3)$

$$\eqalign{\{p_i,\,p_j\}&=0,\cr
\{m_{ij},\,p_k\}=&\,(\delta_{jk}p_i-\delta_{ik}p_j),\cr}\eqn\esixone$$
where $m_{ij}$ are the generators of
 the classical $O(3)$ algebra. We introduce
noncanonical three-coordinates by the relation (see also [19])
$$\{x_i,\,p_j\}=\delta_{ij}\Omega\sp{(\kappa)}( \vec p\sp2).
\eqn\esixonec$$
Then, from (4.3-4) it follows that in the spinless case
$$m_{ij}=(\Omega\sp{(\kappa)}(\vec p\sp2))\sp{-1}
(x_jp_i-x_ip_j).\eqn\efivenew$$
 Moreover, from the Jacobi identity we find that
$$\{x_i,\,x_j\}=M\sp{(\kappa)}( \vec p\sp2)m_{ij},\eqn\esixtwo$$
where, using the relation
$$\{x_i,\,F( \vec p\sp2)\}=2p_i\,\Omega\sp{(\kappa)}( \vec p\sp2)
F'(\vec p\sp2)\eqn\esixthree$$
$M\sp{(\kappa)}$ is given by
$$M\sp{(\kappa)}( \vec p\sp2)={d\over dp\sp2}[\Omega\sp{(\kappa)}(\vec
p\sp2)]\sp2.\eqn\esixfour$$

Moreover, we have
$$\{m_{ij},x_k\}=\delta_{jk}x_i-\delta_{ik}x_j.\eqn\efoursevenb$$

Solutions, with a constant curvature $M\sp{(\kappa)}={1\over 4\kappa\sp2}$
of the three-dimensional phase space are obtained if
$$\Omega\sp{(\kappa)}(\vec p\sp2)=(a+{\vec p\sp2\over 4\kappa\sp2})\sp{1\over
2}.\eqn\esixfive$$
The relation \esixfive\  depends only on the three-momentum but if
 we assume that $p_0$ is given by the \k-relativistic mass-shell
relation (1.2) we can distinguish the following two special cases:
\item{i)}
$a=1+{M\sp2\over 4\kappa\sp2}$. In this case
$$\Omega\sp{(\kappa)}(\vec p\sp2)=cosh\Bigl({p_0\over
2\kappa}\Bigr)\eqn\esixsix$$
and corresponds, after quantisation of the Poisson brackets, to the choice
of the position operator proposed in [\rnineteen].
\item{ii)} We can put $a={M\sp2\over 2\kappa\sp2}$ and obtain
$$\Omega\sp{(\kappa)}(\vec p\sp2)=\sqrt{2}sinh\Bigl({p_0\over
2\kappa}\Bigr).\eqn\esixsixb$$
We note that this case is not very physical as we cannot take
the limit $\kappa\rightarrow\infty$.

The Hamiltonian equations (4.1) corresponding to the  \k-deformed
Poisson brackets (4.3-5)
 take the form
$$\eqalign{\dot x_i=&\Omega\sp{(\kappa)}(\vec p\sp2){\partial H\sp{\kappa}\over
\partial p_i}+M\sp{(\kappa)}(\vec p\sp2)m_{ij}{\partial H\sp{\kappa}\over
\partial x_j},\cr
\dot p_i=&-\Omega\sp{(\kappa)}(\vec p\sp2){\partial H\sp{\kappa}\over \partial
x_i}\cr}\eqn\esixseven$$
and for  the Hamiltonian $H\sp{\kappa}\equiv H_0\sp{\kappa}(\vec p\sp2)$ (see
(1.1)) we get (compare with \efourone)
$$\dot x_i=\Omega\sp{(\kappa)}(\vec p\sp2){\partial H_0\sp{\kappa}\over
\partial
p_i}={\Omega\sp{(\kappa)}(\vec p\sp2)\over \sqrt{1+({\omega\over
2\kappa})\sp2}}{p_i\over \omega}.\eqn\esixeight$$

For $\Omega\sp{(\kappa)}\equiv 1$ we recover the description
with commuting coordinates presented in sect. 3.  Moreover, we see that
 the choice (4.10)  leads to the compensation
of the factor which modifies Einstein's velocity law by a nontrivial
\k-dependent symplectic structure
\ie\ we obtain
$$\dot x_i\sp{M}={p_i\over \omega}.\eqn\efournewthirteen$$
In this case, after quantisation, the operators $x_i$
are non-commuting and can be treated as describing generalised translations
in the curved three-dimensional
 momentum space $S\sp2$ with the radius given by
${1\over \kappa\sp2}$. We see that
 the nonrelativistic momentum space becomes compactified and
 one can expect that the \k-deformation will regularise the
 divergences of the corresponding quantum field theory.
Thus, in this case, the effects of the \k modification of the Hamiltonian
and of the symplectic structure do compensate each other and we
obtain the standard velocity formula (4.13).

Let us add that the relation (4.12) can be derived also by the generalisation
of the formulae (2.7) and (2.9)
as follows $([A,B]_+=AB+BA)$:
$$X_i={1\over 2}[F(P_0),N_i]_+.\eqn\efourfourteena$$
As the equation of motion we can take the Heisenberg relation $\dot
X_i=[X_i,P_0]$ [\rten].
Then, when we restrict our attention to the free \k-relativistic particles, we
recover the equation of motion (4.13) if for $F(P_0)$ in \efourfourteena\
we take
$$F(P_0)={1\over \kappa sinh{P_0\over \kappa}}\Omega\sp{(\kappa)}(\vec
P\sp2)\vert_{\omega=2\kappa\,sinh{P_0\over 2\kappa}}\eqn\efourfifteena$$
and consider the spinless realisations ($\vec P\vec M=0$ in (2.1)).

Finally, let us discuss the freedom in the definition of the \k-deformed energy
operator. We assume that the time translations are generated by the generator
$\tilde P_0$ (see (2.11)) with the nontrivial coproduct (2.14).
Then the equation $\dot Xi=[X_i,\tilde P_0]$ implies, for the commuting
coordinates, the relation
$$\dot X_i\sp{B}={P_i\over \tilde P_0}.\eqn\efoursixteena$$
But when we use $P_i$ and $\tilde P_0$ as our variables the standard
mass-shell relation (2.15) holds and we see that we recover the conventional
Einstein velocity formula.
In this case the \k-deformation is visible only in the modification of the
addition law of energies of independent subsystems.

\chapter{\k-relativistic quantum mechanics}

\noindent \item{a)} Free \k-relativistic particles.

The \k-deformed Schr\"odinger quantum mechanics
of free \k-relativistic particles is described by
a conventional scheme with the Hamiltonian modified according to the
substitution (1.2).
Thus we have
$$i{\partial \over \partial t}\psi(\vec x,t)=H_0\sp{\kappa}({1\over i}\vec
\nabla)\psi(\vec x,t),\eqn\efiveone$$
where $\omega=(\vec p\sp2+m\sp2)\sp{1\over 2}$  and where
$$H_0\sp{\kappa}(\vec p)=2\kappa\,arc\,sinh{\omega\over 2\kappa}=
\omega\,-\,{\omega\sp{3}\over 24\kappa\sp2}\,+\,O({1\over
\kappa\sp4}).\eqn\efivetwo$$
The \k-deformation modifies the energy eigenvalues
of the stationary states as follows:
$$E\,\rightarrow\,E_{\kappa}=2\kappa\,arc\,sinh{E\over
2\kappa}\,=\,2\kappa\,ln\Biggl[{E\over 2\kappa}+\sqrt{{E\sp2\over
4\kappa\sp2}+1}\Biggr].\eqn\efivethree$$
The formula \efivethree, when applied at \k-relativistic energies ($E\gg
\kappa$) leads to the essential modification of the asymptotic behaviour
of the energy spectrum
$$E_{\kappa}\,=\,2\,ln{E\over 2\kappa}\,+\,O({\kappa\over E}).\eqn\efivefour$$

Using the linearity of the \k-deformed Schr\"odinger equation \efiveone\ we can
rewrite
it as
$$iD_t\bigl[{1\over 2\kappa}\bigr]\psi(\vec x,t)=H_0({1\over i}\nabla)\psi(\vec
x,t),\eqn\efiveseven$$
where
$D_t[a]$ denotes the finite difference time derivative with a finite shift in
the purely imaginary direction \ie\
$$\eqalign{\partial_t f(t)\rightarrow&D_t[a]f(t)={1\over
2ai}(f(t+ia)-f(t-ia))\cr
=&{1\over a}sin(a\partial_t)f(t)=\sum_{k=0}\sp{\infty}(-1)\sp{k}{a\sp{2k}\over
(2k+1)!}\partial\sp{2k+1}f(t).\cr}\eqn\efiveeight$$
 From the formal power series expansion we find that
$$2\kappa\,arc\,sinh(iD_t\bigl[{1\over
2\kappa}\bigr])=i\partial_t\eqn\efivenine$$ and so we see that
for any time-independent Hamiltonian $H$  the equations
\efiveone\ and \efiveseven\ are equivalent provided that
$$H_{\kappa}\,=\,2\kappa\,arc\,sinh{H\over 2\kappa}.\eqn\efiveten$$
As the substitution (5.8) is dictated by the \k-relativistic mass-shell
formula (1.2) we call the systems in which this substitution (5.8)
has been made  the \k-relativistic ones.

\noindent \item{b)} Free \k-relativistic oscillator.

The \k-deformed harmonic oscillator is obtained by putting $H={1\over
2}(p\sp2+\omega\sp2x\sp2)$ into \efiveten. The eigenvalues of the \k-deformed
harmonic oscillator   described by the equation
(we set $\hbar=1$)
$$H_{\kappa}\ket{n}=E_n\sp{\kappa}\ket{n}\eqn\efiveten$$
are given by
$$\eqalign{E_n\sp{\kappa}=&2\kappa\,arcsinh{\omega(n+{1\over 2})\over 2\kappa}
\cr
=2\kappa \,ln&\Biggl({\omega(n+{1\over 2})\over
2\kappa}+\sqrt{{\omega\sp2(n+{1\over 2})\sp2\over
4\kappa\sp2}+1}\Biggr).\cr}\eqn\efiveeleven$$
The spectrum \efiveeleven\ has the following limits
$$\lim_{\kappa\rightarrow\infty}\,E_n\sp{\kappa}\,=\,\omega\bigl(n+{1\over
2}\bigr)\eqn\efivetwelve$$
and for very large $n$ ($n\gg \kappa$)
$$E_n\sp{\kappa}=2\kappa\Bigl(ln\bigl(n+{1\over 2}\bigl)-ln{\kappa\over
\omega}\Bigr)+O\bigl({1\over n}\bigr).
\eqn\efivethirteen$$
We see that for large $n$ the spectrum of $H_{\kappa}$ grows as $\kappa\,
ln\,n$.
Assuming that for large distances the harmonic potential energy is more
important (${p\sp2\over 2}+{1\over 2}\omega x\sp2\sim {1\over 2}\omega x\sp2$)
we find that for large $x\sp2$
$$H_{\kappa}=2\kappa \,ln{\omega\sp2 x\sp2\over 2\kappa}\,+\,O({1\over
\kappa\sp2}).\eqn\efivefourteen$$
This result is consistent with the calculations for the three-dimensional
oscillator
with asymptotic logarithmic potential, which provides the asymptotic
logarithmic energy spectrum (see \eg\ [\rtwentyfour]).

Given the \k-deformed spectrum (5.12) we can calculate the partition
function. We find
$$\eqalign{Z_{\kappa}=\sum_{n=0}\sp{\infty}\,exp\{-{E_n\sp{\kappa}\over
kT}\}&=\cr
=\bigl({\omega\over 2\kappa}\bigr)\sp{-{2\kappa\over kT}}\sum_{n=0}\sp{\infty}
[n+{1\over 2}+((n+&{1\over 2})\sp{2}+{4\kappa\sp2\over \omega\sp2})\sp{1\over
2}]\sp{-{2\kappa\over kT}}.\cr}\eqn\ebbbb$$
Introducing the Riemann Zeta function [\rnewone ]
$$\zeta(z,q)=\sum_{n=0}\sp{\infty}{1\over (n+q)\sp{z}},\quad
Re\,z>1\eqn\ezeta$$
we obtain an estimate
$$\zeta({2\kappa\over kT},{1\over 2}+{\kappa\over \omega})\,<\bigl({\omega\over
\kappa}\bigr)\sp{2\kappa\over kT}Z_{\kappa}\,<\zeta({2\kappa\over kT},{1\over
2}).\eqn\ezetaone$$
As $\zeta(z,q)$ has a simple pole at $z=1$ for any finite $q$ we see that the
partition function $Z_{\kappa}$ has a simple pole at
$$T_c={2\kappa\over k}.\eqn\epoleattc$$
Thus we can conclude that the free quantum \k-deformed oscillator describes
a quantum system with a finite Hagedorn temperature $T_c$.
If we define the Helmholz free energy
$$-\beta F=ln\,Z\,\sim\,ln\,(T-T_c)\eqn\efiveeighteen$$
 then the average energy
$$U={\partial \over \partial \beta}(\beta F)=F+\beta {\partial \over \partial
\beta}F\,\sim\,(T-T_c)\sp{-1},\eqn\efivenineteen$$
and the heat capacity
$$C={\partial \over \partial T}U\,\sim\,(T-T_c)\sp{-2}\eqn\efivetwenty$$
we see that the three quantites in (5.18-20)
are all divergent at $T=T_c$. We see that the modes of the free
\k-relativistic oscillators can exist only as a statistical
ensemble in two completely separate phases: $T<T_c$ (the physical
region) and $T>T_c$ (the unphysical one).

\noindent \item{c)} The \k-deformed quantum mechanics.

The two equivalent formulae (5.1) and (5.8) can be also applied to
more general Hamiltonians, namely $H=H_0+V$ with an arbitrary potential $V$.
We see, therefore, that the \k-relativistic Schr\"odinger quantum mechanics can
be
cast in two equivalent forms:
\item{a)} as a conventional quantum mechanics with the modified \k-deformed
Hamiltonian (5.10) (\k-relativistic form).
\item{b)} as a quantum mechanics with a standard Hamiltonian and an imaginary
finite
time difference derivative (see (5.7-8)) (\k-deformed form).

The replacement \efiveeight\ resembles the introduction of the
 finite difference Schr\"odinger equation on the time lattice. We would like
to point out, however, two important distinctions: \hfil\break
\noindent {\sl i)}
 The finite time shift is in the  purely imaginary direction. The equation
\efiveseven\ with  real time shift was introduced
many years ago by Calderola [\rtwentysix] and developed
by Montaldi, Janussis and others (see \eg\ [\rtwentyseven, \rtwentyeight]).
In our scheme the purely imaginary replacement comes from the requirement
that the \k-deformed Poincar\'e algebra is a real Hopf algebra. It appears that
the choice of an imaginary time interval in \efiveeight\
is more suitable from a physical point of view. One can show that
\k-relativistic particles with $sinh\Bigl({p_0\over 2\kappa}\Bigr)$ replaced by
$sin\Bigl({p_0\over 2\kappa}\Bigr)$ in (1.2) (this would correspond to the
introduction of a real lattice in time) attain infinite velocities for
some  finite values of energy.  \hfil\break
\noindent{\sl ii)} We have kept a continuous time coordinate; only the
evolution in time
is expressed by the finite-difference time derivative \efiveeight.

In classical mechanics the formulations with the \k-relativistic Hamiltonian
and with
the replacement \efiveeight\ of the time derivatives describe, in general, two
different deformations; the equivalence holds only for the case of generalised
harmonic oscillators (\ie\ systems with Hamiltonians described
by nondegenerate bilinear forms in phase space).
In sections 3 and 4 we have described only one variant
of a deformed classical mechanics corresponding to the
free \k-modified Hamiltonian (1.2). The \k-deformation of
a general classical mechanics with  time derivatives replaced according to
\efiveeight\  requires  a careful analysis of the problems of associativity
(Jacobi identities) and of the difficulties with obtaining
conserved quantities (in particular energy).
The deformed Poisson brackets with finite difference
derivatives
 will be the subject of a separate publication [\rtwentyfive].

\chapter{Conclusions and outlook}

In this paper we have considered different ways of describing free
\k-relativistic particles. In the classical case we have found two
possibilities:
\item{i)} If we insist that the energy is additive we can have either commuting
space coordinates and a strange velocity formulae (velocity going to zero at
infinite energy) or noncommuting coordinates and the standard Einstein velocity
formula.
\item{ii)} If we are prepared to accept a nonlinear addition law for the
 energy, \ie\ use formulae (2.13) describing our \k-Poincar\'e algebra, then we
 can have the commuting space coordinates as well as the standard Einstein
velocity formula.

Let us add, however, that in this paper we have not used in any way the
non-cocommutative coproduct for the three-momenta, which can be
interpreted as the following addition law
$$p_i\sp{(1+2)}=p_i\sp{(1)}\,exp\{{-{p_0\sp{(2)}\over 2\kappa}}\}+
p_i\sp{(2)}\,exp\{{-{p_0\sp{(1)}\over 2\kappa}}\}.\eqn\efinalone$$
The nonstandard coproducts of the four-momentum generators imply the
noncommutativity
of the
generators $\hat x_i,\,\hat x_0$ of the dual quantum Poincar\'e group described
by the relations (1.6). In consequence, the commutativity
of the four-momentum  generators imply that the coproduct of the quantum space
translation generators $\hat x_{\mu}$
is classical
$$\Delta(\hat x_{\mu})=\hat x_{\mu}\otimes 1+1\otimes \hat
x_{\mu},\eqn\efinaltwoa$$
or, equivalently,  we obtain the classical addition formula for the eigenvalues
of $\hat x_{\mu}$:
$$x_{\mu}\sp{(1+2)}=x_{\mu}\sp{(1)}+x_{\mu}\sp{(2)}.\eqn\efinaltwob$$

The duality relation between the \k-deformed Hopf algebra of the four-\break
momentum generators (commutative-noncocommutative) $H_p$ and the Hopf algebra
of the quantum four-translations (noncommutative-cocommutative) $H_x$ implies
the modification of the classical Fourier transform relating momentum and
 coordinate spaces.
With a quantum double $H_p\otimes H_x$ we can define on it two \k-deformed
exponential functions (called $T$ matrix in [\rtwentynine]; see also [22,
\rthirty]) with the properties:
$$\eqalign{e\sp{ix_{\mu}p\sp{\mu(1)}}e\sp{ix_{\mu}p\sp{\mu(2)}}&=e\sp{ix_{\mu}p\sp{\mu(1+2)}}\cr
e\sp{ix_{\mu}\sp{(1)}p\sp{\mu}}e\sp{ix_{\mu}\sp{(2)}p\sp{\mu}}&=
e\sp{ix_{\mu}\sp{(1+2)}p\sp{\mu}}.\cr}\eqn\efinalthree$$

It is clear that the quantum space picture consistent with the Hopf algebra
structure of the \k-deformed Poincar\'e algebra should be obtained
by performing
the \k-deformed Fourier transforms from the commuting four-momentum space to
the noncommuting space coordinates (1.6) using the transformation kernels
satisfying \efinalthree. At present it is not clear to the authors how the
noncommuting space coordinates $x_{\mu}$ and non-additive four-momenta
variables $p\sp{\mu}$ can be incorporated into  consistent \k-deformed schemes
of classical and quantum mechanics.

\endpage
\ack
We would like to thank H. Bacry,
M. Dubois-Violette,  D.B. Fairlie, C. Gomez,  H. Grosse, J. Madore
and M. Ruiz-Altaba for discussions and valuable comments.
One of us (JL) would like to thank the EC for a grant which has
made his visit to Durham possible.
\refout
\endpage
\bye